\documentclass[manuscript]{aastex}
\usepackage{amsbsy,amssymb,amsmath,bm}
\usepackage{graphicx,subfigure}
\usepackage{tabularx}
\usepackage{natbib}

\usepackage{color}  
\newcommand{\unit}[1]{{\rm\,#1}}
\renewcommand{\Re}{\textit{Re}}
\newcommand{\Rm}{\textit{Rm}}
\newcommand{\Va}{\textit{Va}}
\newcommand{\VA}{V_{\rm A}}
\newcommand{\Pm}{\textit{Pm}}
\newcommand{\Rsun}{\mathrm{R_\odot}}

\begin{document}

\title{On obliquely magnetized and differentially rotating stars}
\author{Xing Wei, Jeremy Goodman}
\affil{Princeton University Observatory, Princeton N.J. 08544 USA}
\email{xingwei@astro.princeton.edu, jeremy@astro.princeton.edu}

\begin{abstract}
We investigate the interaction of differential rotation and a misaligned magnetic field.  The incompressible magnetohydrodynamic equations are solved numerically for a free-decay
problem.  In the kinematic limit, differential rotation annihilates the non-axisymmetric field on a timescale proportional to the cube root of magnetic Reynolds number ($\Rm$), as predicted by R\"adler. Nonlinearly, the outcome depends upon the initial energy in the non-axisymmetric part of the field.   Sufficiently weak fields approach axisymmetry as in the kinematic limit; some differential rotation survives across magnetic surfaces, at least on intermediate timescales. Stronger fields enforce uniform rotation and remain non-axisymmetric.  The initial field strength that divides these two regimes does not follow the scaling $\Rm^{-1/3}$ predicted by quasi-kinematic arguments, perhaps because our $\Rm$ is never sufficiently large or because of
reconnection.  We discuss the possible relevance of these results to tidal synchronization and tidal heating of close binary stars, particularly double white dwarfs.
\end{abstract}

\maketitle

\section{Introduction}\label{sec:intro}

Current understanding of stellar differential rotation leaves much to be desired.  Centuries of sunspot observations show that the Sun rotates more quickly than its equator.  Helioseismology reveals that this latitudinal variation extends throughout the convection zone, but contrary to expectations for a (nearly) isentropic region, the angular velocity is not constant on cylinders \citep{Schou+etal1998}.  There is no consensus as to how this pattern is maintained, though the convection itself is presumably essential.  Inversions for the radiative core are consistent with uniform rotation down to at least $0.2 R_\odot$ \citep{Chaplin+etal1999}; constraints at greater depth are weak because they depend on a few low-degree p modes.  Asteroseismological analyses of {\it Kepler} photometry find that many giants and subgiants have cores that rotate more rapidly than their envelopes, but not so rapidly as if these cores had contracted at constant angular momentum \citep{Deheuvels+etal2012,Mosser+etal2012,Deheuvels+etal2014}.  Magnetic transport of angular momentum is likely responsible \citep{Maeder+Meynet2014}, but standard prescriptions used by the stellar-evolution community fail to explain the observations quantitatively \citep{Eggenberger+etal2012,Cantiello+etal2014}.  White dwarfs, the end-states of low-mass stars, are known to rotate slowly, with periods ranging from hours to days (\citealt{Berger+etal2005,Kawaler2014} and references therein).  Asteroseismological attempts to measure \emph{differential} rotation in the interiors of pulsating white dwarfs have so far produced ambiguous results \citep{Charpinet+etal2009,Corsico+etal2011}. The role of stellar magnetic fields in the transport of angular
momentum in all of these objects may be crucial but is not yet well understood. The present work addresses one aspect of
this complex problem.

Ferraro's isorotation law applied to stellar interior states that ``the star can possess a steady field only if the field is symmetric about the axis of rotation, and each line of force lies wholly in a surface symmetric about the axis and rotating with uniform angular velocity'' \citep{Ferraro}.  A detailed derivation can be found in \cite{Cowling}. The theorem can be mathematically expressed as $\bm B_p\cdot\bm\nabla\Omega=0$, where $\bm B_p$ is the axisymmetric meridional field and $\Omega$ the angular velocity. This formula shows that the contours of axisymmetric meridional field are parallel to the contours of angular velocity. The isorotation law is strictly valid only for a perfect conductor, i.e. the magnetic Reynolds number $\Rm\rightarrow\infty$. Later, \citet{Mestel_Weiss} studied the dynamical and resistive effects of departures from the isorotation law. Their results suggest that Alfv\'en waves transfer angular momentum along meridional field lines---at different speeds on different lines---until the isorotation law is achieved. This mechanism of ``phase-mixing'' was studied by \citet{Ionson1978,Heyvaerts+Priest1983,Spruit1999}. The latter estimated that the wave amplitude should decay as $e^{-(t/t_p)^3}$ on a phase-mixing timescale $t_p\sim (R^4/\eta V^2_{\textsc{a}})^{1/3}$, $R$ being stellar radius, $\eta$ being magnetic diffusivity, and $V_{\textsc{a}}$ being a typical Alfv\'en speed. Therefore, it may be inferred that if all field lines are attached to a uniformly rotating solid core, then the entire stellar (or planetary) interior will eventually achieve solid-body rotation (e.g., \citealt{Charbonneau_MacGregor}).  \citet{Mestel_Weiss} suggested that even without such a core, a fluid body will tend toward solid-body rotation if its magnetic field is significantly non-axisymmetric: for example, a dipolar field whose axis is not parallel to the rotation axis, a situation referred to in our paper as an \emph{oblique rotator}. It is this last assertion, rather than the whole subject of stellar differential rotation, that is the main focus of the present paper.

Mestel and Weiss perhaps overstated their argument by restricting the velocity field to pure rotation, rather than
a combination of rotation and meridional circulation.  Consider, for example, a star in a nontrivial state of isorotation aligned with its
magnetic axis, but undergoing slow precession around another axis (due, perhaps, to the tidal torque of a companion). Such a star
would not be strictly axisymmetric with respect to the axis defined by its total angular momentum, but its velocity field---which would
have meridional as well as azimuthal components with respect to that axis---could satisfy Ferraro's Law and be steady in a frame rotating  with the star's body axes.  The calculations described in this paper allow for meridional motions.

Even as Mestel and Weiss posed the problem, the
outcome surely depends on the relative strengths of the field and of differential rotation. \citet{raedler} considered the kinematic limit in which the backreaction of the magnetic field on the flow is neglected, so that the field evolves initially linearly according to the induction equation in a prescribed flow.  He concluded that the combined effects of differential rotation and magnetic diffusion cause the non-axisymmetric field to decay more rapidly than the axisymmetric field, thus tending to reduce the magnetic obliquity.  Allowing for magnetic forces, \citet[ \S9.3]{Mestel}  concluded that a nonaxisymmetric meridional field with a weak component perpendicular to the rotational axis can be destroyed by differential rotation, but a stronger perpendicular component can destroy the differential rotation.  The principal goal of the present paper is to test and quantify these conclusions by explicit calculations.

The problem posed here has several applications in geophysics and astrophysics.  In many stars and fluid planets, it will often be complicated by a magnetic dynamo, which may reinforce the magnetic field in ways that cannot confidently be predicted.  We are motivated, however, mainly by applications to white dwarfs, whose magnetic fields are probably relics inherited from their progenitors.  
The interaction between differential rotation and magnetic fields may therefore be simpler to study in the context of white dwarfs
than in main-sequence or red-giant stars.  Admittedly the observational constraints are weaker in white dwarfs.  On the other hand,
the consequences might be more dramatic, as we now explain.

In the double-degenerate scenario for Type Ia supernovae (SNe Ia), two white dwarfs orbiting one another are gradually driven together by gravitational radiation.  At late phases of the inspiral when the stars are separated by a modest multiple of their radii, they will exert mutual tidal torques tending to enforce synchronism between their rotational and orbital frequencies.  These torques, which likely involve resonant excitation of inertial oscillations and internal waves (g modes), are expected to be unequally distributed within the star, in fact concentrated toward the surface where thermal timescales are shortest, densities are least, and tidally excited waves may break nonlinearly \citep{Fuller+Lai2012,Burkart+etal2013,Dall'Osso+Rossi2014}, possibly with observable consequences \citep{Fuller+Lai2013}.
When the star rotates rapidly compared to the tidal frequency, i.e. when $\Omega_{\rm spin}\gg\Omega_{\rm orbit}-\Omega_{\rm spin}$, the torque is also concentrated in latitude, toward the equator \citep{Fuller+Lai2014}.
Magnetic stresses are likely needed to couple the rotation of the stellar interior to that of the surface layers.  The amount of associated dissipation in the interior may depend not only on the strength of the magnetic field and of the tide, but also on the symmetry of the field.  If it is essentially axisymmetric, then since plasma viscosity is likely negligible, some turbulent dissipation is probably required to transport angular momentum across the lines.  

Also, the efficiency of magnetic redistribution of angular momentum may affect the degree of nonsynchronous rotation, which, even if small, determines the total dissipation associated with a given tidal torque.  An upper bound to the tidal heating rate is the power required to maintain synchronous rotation, $\dot E_{\rm spin}= (I_1+I_2)\Omega\dot\Omega$, where $\Omega$ is the orbital angular velocity and $I_{1,2}$ are the moments of inertia of the two stars.  For two $0.7\,M_\odot$ carbon-oxygen white dwarfs driven together by gravitational radiation, $\dot E_{\rm spin}\approx 10^{38} P_{\rm min}^{-14/3} \unit{erg\,s^{-1}}$, where $P_{\rm min}$ is the orbital period in minutes, while the time before contact is $400 P_{\rm min}^{-8/3}\unit{yr}$.  As noted in the works cited above, however, the actual dissipation rate will be less than this by an appropriate average of $(1- \Omega_{\rm spin}/\Omega_{\rm orb})$, $\Omega_{\rm spin}$ being the rotational angular velocity of each mass element (which differs among elements if the star rotates differentially).  Thus the dissipation will be quite small if the tidal torques are efficiently redistributed so that all parts of the star are kept nearly synchronous with the orbit.  The dissipation will also be small if the tidal torques are weak so that $\dot\Omega_{\rm spin}\ll\dot\Omega_{\rm orb}$.  Still, if even a small fraction of $\dot E_{\rm spin}$ were dissipated and the heat were transported to the surface, such systems could be quite luminous.

Surface magnetic fields of white dwarfs vary widely.  Zeeman measurements indicate that $\sim10\%$ of these stars have fields exceeding $2$~MG \citep{Liebert+etal2003}, but sensitive polarimetry (of admittedly small samples) suggest that the majority have surface fields $\lesssim10\unit{kG}$ \citep{Landstreet+etal2012}.  In the expected absence of dynamo action, electrical conductivities in the degenerate interiors of these stars are such that their fields should decay on timescales $\sim 10^9\unit{yr}$ \citep{Fontaine+etal1973}.  This is of the same order as the ages of most observed white dwarfs and likely also of many SN~Ia progenitors.  Depending upon the star's mass, crystallization begins at the center after one to a few billion years when the luminosity has fallen to $10^{-3}$ to $10^{-4}\unit{L_\odot}$ \citep{Renedo+etal2010}.  Some field lines may be anchored to the growing solid core, an effect not accounted for in the idealized models considered here.

The double-degenerate scenario for SNe~Ia, though favored by recent evidence, is unproven \citep{Maoz+etal2014}.  A variant invokes merging via head-on collisions in a triple-star system rather than inspiral driven by gravitational waves \citep{Katz+Dong2012,Kushnir+etal2013}.  Yet short-period white-dwarf binaries do exist \citep{Nelemans+etal2004,Brown+etal2011}, and their end states invite speculation even if they are not destined to be supernovae.

The plan of this paper is as follows.  \S2 frames the free-decay problem and our numerical methods.  \S3 compares results for the kinematic problem, in which the differential rotation is prescribed and the backreaction of the field neglected, with the predictions of \cite{raedler}.  \S4 presents results for the full nonlinear problem, including estimates for the critical initial field strength as a function of $\Rm$.  Numerical considerations limit our calculations to $\Re=\Rm\le 10^4$, much less than in a real white dwarf.  Finally, \S5 discusses the relationship of these numerical results to the astrophysical problem that motivates this work.  It is clear that there are many aspects of the joint tidally-driven evolution of the magnetic field and differential rotation that will need further study.

\section{Formulation}\label{sec:formulation}

We calculate the three-dimensional and fully nonlinear unforced MHD equations in a spherical shell with inner radius $r_i$ and outer radius $r_o$. The differential rotation and misaligned field are given as initial conditions. To conserve total angular momentum, stress-free conditions are imposed on the fluid velocity and insulating  conditions on the magnetic field at both boundaries.

The dimensionless MHD equations for an incompressible fluid of constant and uniform density read
\begin{equation}\label{ns}
\frac{\partial\bm u}{\partial t}+\bm u\cdot\bm\nabla\bm u=-\bm\nabla p +\frac{1}{\Re}\nabla^2\bm u+(\bm\nabla\times\bm B)\times\bm B,
\end{equation}
\begin{equation}\label{induction}
\frac{\partial\bm B}{\partial t}=\bm\nabla\times\left(\bm u\times\bm B\right)+\frac{1}{\Rm}\nabla^2\bm B.
\end{equation}
Length has been normalized to $r_o$, time to $\Omega_0^{-1}$, velocity to $\Omega_0r_o$ where $\Omega_0$ is characteristic of the initial angular velocity [see eq.~\eqref{initialflow}], pressure to $\rho\Omega_0^2r_o^2$, and magnetic field to $\sqrt{\rho\mu}\Omega_0r_o$. There are three dimensionless parameters governing the MHD flow. The Reynolds number $\Re=\Omega_0r_o^2/\nu$, where $\nu$ is kinematic viscosity, is the ratio of the viscous time scale to the initial rotational time scale. Similarly, the magnetic Reynolds number $\Rm=\Omega_0r_o^2/\eta$, where $\eta$ is magnetic diffusivity, is the timescale for diffusion of the initial magnetic field relative to the rotation time. For numerical feasibility, we keep $\Re=\Rm$, namely the magnetic Prandtl number $\Pm$ is unity. The dimensionless Alfv\'en velocity $\Va=B_0/(\sqrt{\rho\mu}\,\Omega_0r_o)$, where $B_0$ is characteristic of the initial field [see eq.~\eqref{initialfield}], measures the strength of initial field relative to initial rotation. To minimize the effect of the inner sphere while avoiding the coordinate singularity at the origin, we take $r_i/r_0=0.1$.

The details of numerical method can be found in \cite{Hollerbach}.  Toroidal-poloidal decompositions
 are used to guarantee $\bm\nabla\cdot\bm u=\bm\nabla\cdot\bm B=0$:
\begin{equation}
\bm u=\bm\nabla\times(e\,\hat{\bm e}_r)+\bm\nabla\times\bm\nabla\times(f\,\hat{\bm e}_r), \quad \bm B=\bm\nabla\times(g\,\hat{\bm e}_r)+\bm\nabla\times\bm\nabla\times(h\,\hat{\bm e}_r).
\end{equation}
Note that the toroidal part $\bm\nabla\times(e\,\hat{\bm e}_r)$ has a latitudinal component proportional to
$\partial e/\partial\phi$, and that the poloidal part $\bm\nabla\times\bm\nabla\times(f\,\hat{\bm e}_r)$ has an
azimuthal component proportional to $\partial^2 f/\partial\phi\partial r$.  Therefore, when speaking of non-axisymmetric
fields, we will refer to components parallel to ${\bm e}_\phi$ as \emph{azimuthal} (rather than toroidal), and 
to components parallel to the $r\theta$ plane as \emph{meridional} (rather than poloidal),
to avoid confusion caused by experience with axisymmetry, where
``toroidal'' is synonymous with ``azimuthal,'' and ``poloidal'' with ``meridional.''

In spherical coordinates $(r,\theta,\phi)$ the functions $\{e,f,g,h\}$ are expanded in the angular coordinates with spherical harmonics and in radius with Chebyshev polynomials.  For example,
\begin{align}\label{expansion}
&e(r,\theta,\phi,t)=\sum_{l,m}\left[e^c_{lm}(r,t)\,P_l^m(\cos\theta)\,\cos(m\phi)+e^s_{lm}(r,t)\,P_l^m(\cos\theta)\,\sin(m\phi)\right], \nonumber\\
&e^c_{lm}(r,t)=\sum_k e^c_{klm}(t)\,T_k(x) \mbox{ and } e^s_{lm}(r,t)=\sum_k e^s_{klm}(t)\,T_k(x),
\end{align}
where $x=(2r-r_o-r_i)/(r_o-r_i)\in[-1,+1]$.  We use a second order Runge-Kutta scheme for time stepping.  The diffusive terms are treated implicitly.

We impose stress-free boundary condition for fluid velocity at both $r_o$ and $r_i$, namely $ u_r=\tau_{r\theta}=\tau_{r\phi}=0$. Translated to spherical harmonics, this becomes
\begin{equation}\label{u_bc}
f_{lm}=\ \frac{d}{dr}\left(\frac{1}{r^2}\frac{d}{dr}f_{lm}\right)=\ \frac{d}{dr}\left(\frac{1}{r^2}e_{lm}\right)=0\,.
\end{equation}
These hold for both cosine and sine components and so the superscripts $c$ and $s$ are omitted. We require the field to match onto potential fields interior to $r_i$ and exterior to $r_o$ that are regular at $r=0$ and $r=\infty$, respectively, so that
\begin{equation}\label{b_bc}
g_{lm}=\ \left(\frac{d}{dr}-\frac{l+1}{r}\right)h_{lm}=0 \mbox{ at } r_i, \quad
g_{lm}=\  \left(\frac{d}{dr}+\frac{l}{r}\right)h_{lm}=0 \mbox{ at } r_o.
\end{equation}

The initial differential rotation profile is taken for the hydrostatic equilibrium, i.e. inertial force is balanced by pressure gradient, such that the angular velocity depends only on cylindrical radius $R=r\sin\theta$,
\begin{equation}\label{initialflow}
\Omega=\Omega_0r^2\sin^2\theta.
\end{equation}
Nonmagnetic force balance with more general patterns of differential rotation would require stratification, which
  we wish to avoid.  It seems unlikely that this simplification qualitatively affects the competition between differential rotation
and non-axisymmetry, but it does restrict the allowable isorotational states in case the former triumphs over the latter.
A larger exponent of $R$ might better imitate the concentration of tidal torques toward the surface and equator but would lead to rapid magnetic and viscous diffusion at the numerically accessible values of $\Re$ and $\Rm$.  Like this profile,
the differential rotation resulting from tidal torques in an inspiraling binary is probably immune to magnetorotational instability (MRI) because $\partial\Omega^2/\partial R>0$ (e.g., \citealt{Balbus2003}).

\begin{figure}[h]
\centering
\subfigure[]{\includegraphics[scale=0.2]{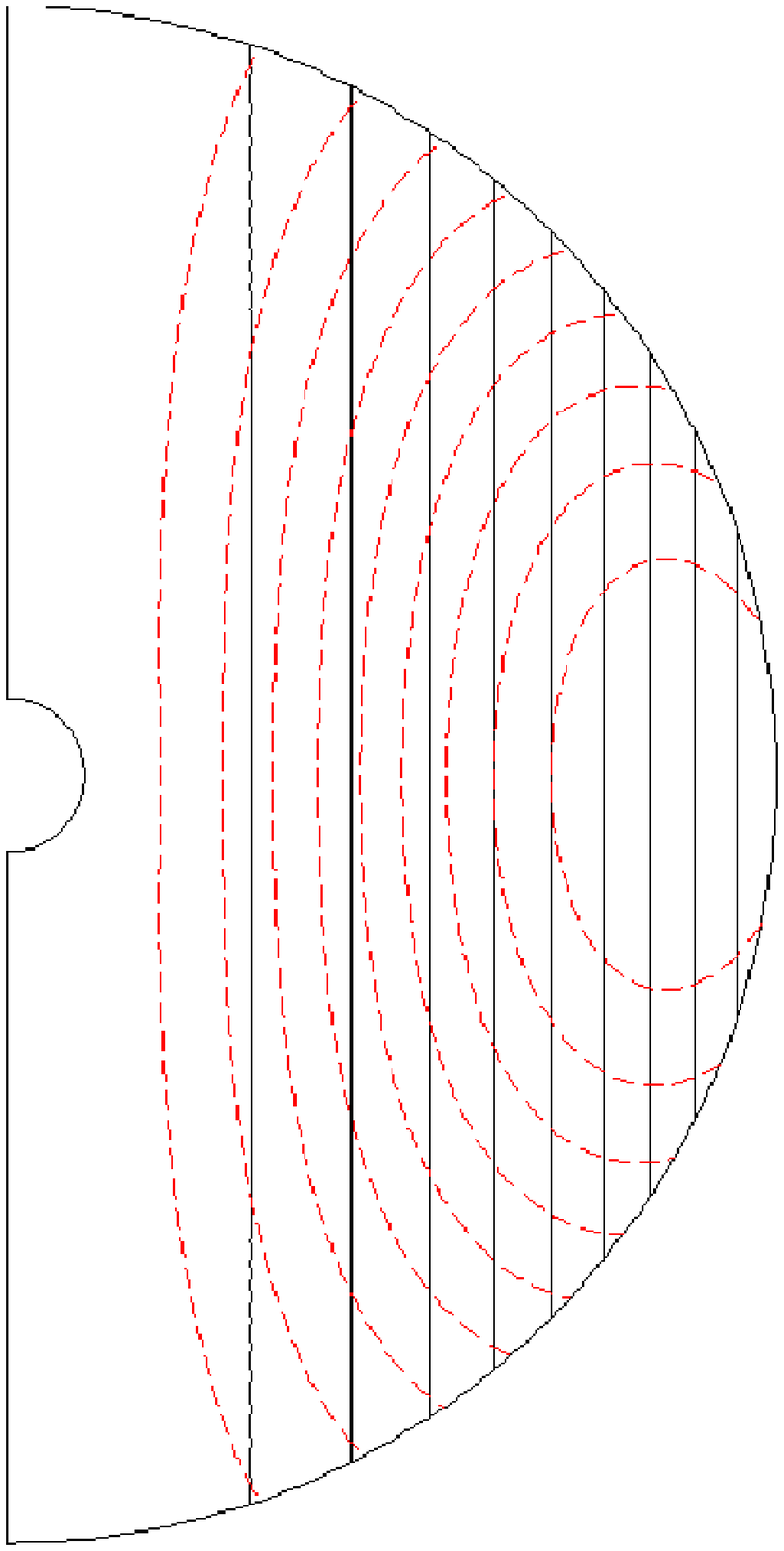}\label{b0-omega0}}
\subfigure[]{\includegraphics[scale=0.7]{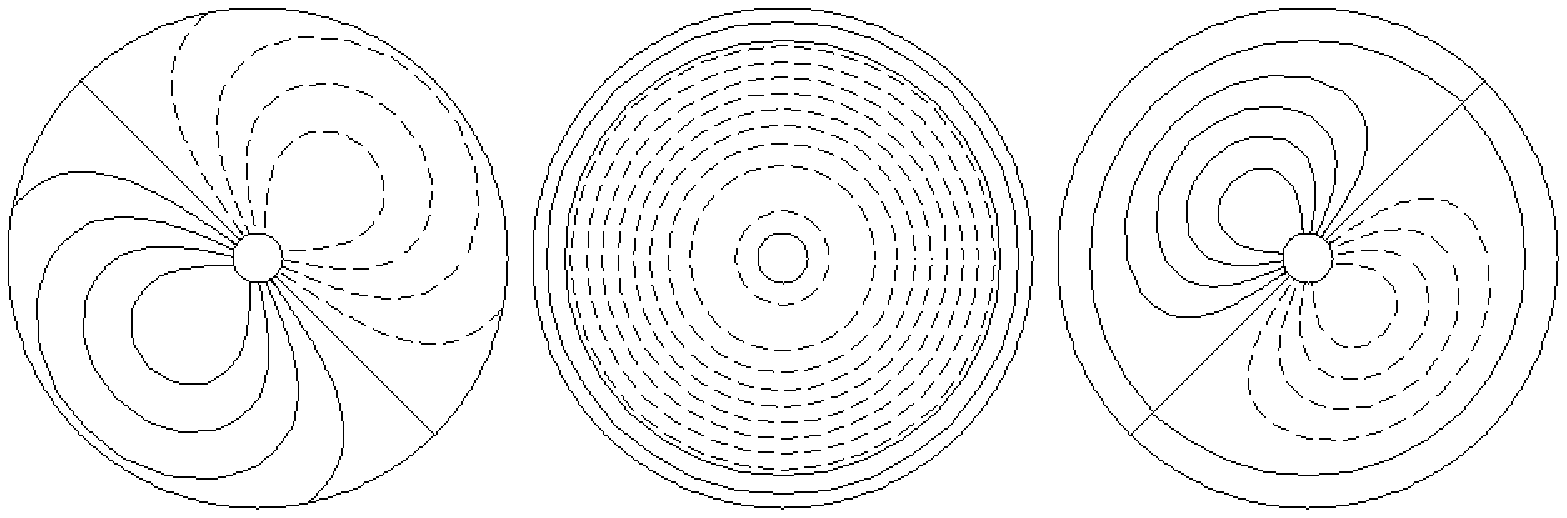}\label{b0}}
\caption{Initial conditions with magnetic obliquity $\alpha=45^\circ$.
(a) Contours of initial axisymmetric angular velocity (black lines) and poloidal field (red lines) in the meridional plane. (b) Contours of the components of initial field in the equatorial plane: $B_r$, $B_\theta$ and $B_\phi$ (left to right). In both (a) and (b), solid lines denote positive and dashed negative.}\label{initial}
\end{figure}

For the initial field we choose free-decay modes, eigenfunctions of the induction equation with finite conductivity and currents confined to the body (\citealt{Moffatt}, \S2.7). If the conductivity is uniform, the poloidal expansion coefficients obey
\begin{equation}
\frac{\partial h_{lm}}{\partial t}=\frac{\partial^2h_{lm}}{\partial r^2}-\frac{l(l+1)}{r^2}h_{lm}\,.
\end{equation}
The eigenfunctions for $h_{lm}$ are linear combinations of spherical Bessel functions of the first and second kind that satisfy the insulating boundary condition \eqref{b_bc}. We choose for the initial conditions a linear combination of the two lowest-order poloidal free-decay modes $(l=1,m=0)$ and $(l=1,m=1)$:\footnote{These modes suffice to match onto a general dipole field in the vacuum exterior to the star. In white dwarfs, modes with higher $l$ and/or more radial nodes would undergo significant resistive decay on Gyr timescales.}
\begin{equation}
h(r)=h_{10}(r)\cos\alpha+h_{11}(r)\sin\alpha,
\end{equation}
in which $\alpha$ is the angle between the rotational and magnetic axes. The initial field is then
\begin{align}\label{initialfield}
B_r&=B_0\frac{2}{r^2}\left[h^c_{10}\cos\theta\cos\alpha-\sin\theta\left(h^c_{11}\cos\phi+h^s_{11}\sin\phi\right)\sin\alpha\right], \nonumber\\
B_\theta&=-B_0\frac{1}{r}\left[\frac{dh^c_{10}}{dr}\sin\theta\cos\alpha+\cos\theta\left(\frac{dh^c_{11}}{dr}\cos\phi+\frac{dh^s_{11}}{dr}\sin\phi\right)\sin\alpha\right], \nonumber\\
B_\phi&=B_0\frac{1}{r}\left(\frac{dh^c_{11}}{dr}\sin\phi-\frac{dh^s_{11}}{dr}\cos\phi\right)\sin\alpha\,.
\end{align}
The normalization is chosen so that the initial magnetic energy is $\Va^2\rho\Omega_0^2r_0^5$, where $\rho\Omega_0^2r_0^5$ is unity in our scaled units. The initial field has a nonzero azimuthal component unless $\sin\alpha=0$. Figure \ref{b0-omega0} shows the meridional distributions of initial angular velocity (black lines) and poloidal field (red lines) and figure \ref{b0} shows the three components of initial field in the equatorial plane for $\alpha=45^\circ$.

With the boundary conditions, the total angular momentum should be conserved, and with our initial conditions, it is purely axial:
\begin{equation}
L_z=\int_V r\sin\theta\,u_\phi dV=\frac{4}{3}\pi(r_o-r_i)\sum_ke^c_{k10}\int_{-1}^{+1}r^2T_k(x)dx.
\end{equation}
The radial integral can be found analytically. In our numerical calculations, $e^c_{k10}$ is monitored at each time step, and the total angular momentum is found to be conserved to one part in $10^8$. For the initial rotation profile \eqref{initialflow}, $L_z=0.9574$ in our units, and the initial kinetic energy $E_0=0.3191$.  Because of viscosity, uniform (solid-body) rotation must eventually be established.  The final angular velocity and kinetic energy are $\Omega_\infty=L_z/I=0.5714$ and $E_\infty = L_z^2/2I=0.2736$, where $I\approx 1.676$ is the moment of inertia of the fluid shell. Therefore, the excess kinetic energy available for dissipation is $E_0-E_\infty\approx0.0455\approx E_0/7$.  In \S4, we compare the times required for the magnetic and \emph{excess} kinetic energies to decay to 10\% of their initial values, as functions of $\Rm=\Re$ and $\Va$.

\section{Kinematic problem}\label{sec:kinematic}

Before addressing the fully nonlinear problem, we study a simplified kinematic one in which the flow is fixed in its initial
form \eqref{initialflow}, and only the magnetic induction equation \eqref{induction} is solved. This is a linear problem and the spherical harmonics decouple. Therefore, the axisymmetric and non-axisymmetric fields evolve independently.

\citet{raedler}'s analysis applies here, and we summarize it.
Differential rotation shears the axisymmetric meridional field into axisymmetric azimuthal field---the so-called $\omega$ effect---with the same dependence on $(r,\theta)$ and therefore a constant characteristic decay time $t_d=r_o^2/\eta$. However,
if diffusion is neglected, 
the non-axisymmetric field reverses on progressively finer lengthscales as time goes on.  R\"adler demonstrates this with a
cartoon of a field line winding up in a plane, but the point is important enough to us that we give a more careful argument.
In ideal MHD, advection by the velocity field $\Omega(r,\theta) r\sin\theta\,\hat{\bm e}_\phi$ preserves the meridional components $\bm B_p\equiv (B_r,B_\theta)$ along the flow, as can be seen by considering that any  closed fluid contour drawn on a sphere $r=$constant remains on the sphere and encloses constant area and constant flux as it is advected; and similarly for any contour on a cone $\theta=$constant. Therefore, if $\bm B_p^{(t)}(r,\theta,\phi)$ is the meridional field at time $t$, then the evolution of this field and its derivatives is
\begin{align*}
\bm B_p^{(t)}(r,\theta,\phi) &= \bm B_p^{(0)}[r,\theta,\phi-\Omega(r,\theta)t]\,,\\
\frac{\partial}{\partial r}\bm B_p^{(t)}(r,\theta,\phi) &=
\frac{\partial}{\partial r}\bm B_p^{(0)}[r,\theta,\phi-\Omega(r,\theta)t]\ +t\,\frac{\partial\Omega}{\partial r}
\frac{\partial}{\partial \phi}\bm B_p^{(0)}[r,\theta,\phi-\Omega(r,\theta)t]\,,\\
\frac{\partial}{\partial\theta}\bm B_p^{(t)}(r,\theta,\phi) &=
\frac{\partial}{\partial\theta}\bm B_p^{(0)}[r,\theta,\phi-\Omega(r,\theta)t]\ +t\,\frac{\partial\Omega}{\partial\theta}
\frac{\partial}{\partial \phi}\bm B_p^{(0)}[r,\theta,\phi-\Omega(r,\theta)t]\,,
\end{align*}
from which it can be seen that the meridional derivatives
$\partial_r\bm B_p$ and $\partial_\theta\bm B_p$ increase linearly with time unless $\partial\bm B_p^{(0)}/\partial\phi=0$ (axisymmetry) or $\bm\nabla\Omega=0$ (solid-body rotation).  Yet the first line above implies that $\int |\bm B^{(t)}_p|^2dV$ is constant.  Therefore, the non-axisymmetric part of $\bm B_p^{(t)}$ must reverse direction on progressively smaller scales, as was to be shown.  Finally, since the meridional field is the source of the growing azimuthal field via the term $\propto \bm B_p\cdot\bm\nabla\Omega$ in the induction equation, the non-axisymmetric part of $B^{(t)}_\phi$ must also develop such reversals.

Given a small diffusivity $\eta$, it follows that the resistive timescale of the $m^{\rm th}$ azimuthal harmonic decreases as $t_{d,m} \sim \eta^{-1} m^{-2}|\bm\nabla\Omega|^{-2}\,t^{-2}$ at late times ($m\ne0$) where the length scale is taken to be $(m|\bm\nabla\Omega|t)^{-1}$, i.e. the length scale of shear $(|\bm\nabla\Omega|t)^{-1}$ divided by number of polarity reversals $m$. Once $t_{d,m}\lesssim t$, all components of the $m^{\rm th}$ harmonic quickly decay.
Thus the toroidal field and magnetic energy will peak at a time
\begin{equation}
\label{eq:tpeak}
t_{\rm peak} \sim
\begin{cases} 
\left(m^{-2}\eta^{-1}|\bm\nabla\Omega|^{-2}\right)^{1/3}\equiv m^{-2/3} (\Rm')^{1/3}(\Delta\Omega)^{-1} & m\ne0,\\
\phantom{m^{-2}\eta^{-1}|\bm\nabla\Omega|^{-2}\equiv m^{-2/3}\qquad} \Rm'\,(\Delta\Omega)^{-1} & m=0\,,
\end{cases}
\end{equation}
where $\Delta\Omega\sim r_0|\bm\nabla\Omega|_{\rm rms}$ is a measure of the differential rotation, and $\Rm'\equiv r_0^2\Delta\Omega/\eta$.
The peak magnetic energy density therefore scales as
\begin{equation}
\label{eq:Epeak}
E_{\rm mag,\,peak} \sim
\begin{cases}
(\Rm')^{2/3} E_{\rm mag,\,init} m^{-4/3} & m\ne0,\\
(\Rm')^2 E_{\rm mag,\,init}  & m=0,
\end{cases}
\end{equation}
unless the $m=0$ component of the field is isorotational.

\begin{figure}[h]
\centering
\includegraphics[scale=0.5]{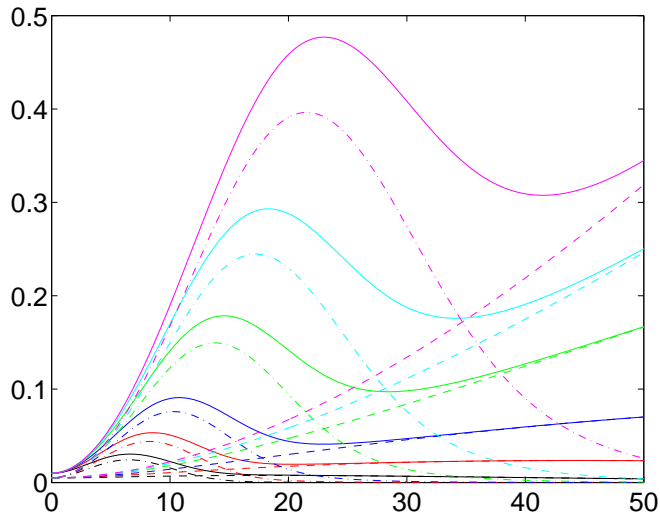}
\caption{Time evolution of total (solid), axisymmetric (dashed) and non-axisymmetric (dash-dot) magnetic energy in the kinematic problem at $\Va=0.1$ and $\alpha=45^\circ$. Curves from bottom to top (colored black, red, blue, green, cyan and magenta) denote respectively $\Rm=500$, $1000$, $2000$, $5000$, $10000$ and $20000$.}\label{kinematic}
\end{figure}

Figure \ref{kinematic} shows the time evolution of total, axisymmetric and non-axisymmetric magnetic energy at different
$\Rm$ with $\Va=0.1$ and $\alpha=45^\circ$. Initially, the axisymmetric azimuthal magnetic field grows linearly with
time and its energy quadratically, but eventually all components of the field decay resistively. The non-axisymmetric
energy grows faster and decays much faster than the axisymmetric energy. The peak value of non-axisymmetric magnetic
energy and the time at which it is achieved are given in Table~\ref{scale}. These results can be fit by
$t_{\rm peak}\propto \Rm^{0.323\pm0.004}$ and $E_{\rm mag,\, peak}\propto \Rm^{0.753\pm0.035}$, in rough agreement with
equations \eqref{eq:tpeak} and \eqref{eq:Epeak}. The exponent of the latter fit is influenced by behavior at
small $Rm$; between the two highest-$Rm$ points, the logarithmic slope is $0.696$. A toy model that we will not go
into here suggests that both exponents can be expected to differ from their asymptotic values as $Rm\to\infty$ by
corrections of relative size $\sim Rm^{-1/3}$. This explains the deviations of $E_{\rm mag,\,peak}$ from its expected
asymptotic scaling rather well, leaving as a mystery why the $t_{\rm peak}$ scaling does not deviate more than
observed.

\begin{table}
\begin{tabularx}{\textwidth}{X|X|X|X|X|X|X}
$\Rm$ & 500 & 1000 & 2000 & 5000 & 10000 & 20000 \\
\hline
$t_{\rm peak}$ & 6.5 & 8.2 & 10.3 & 13.8 & 17.2 & 21.5 \\
\hline
$E_{\rm mag,\,peak}$ & 2.431E-2 & 4.396E-2 & 7.591E-2 & 1.495E-1 & 2.447E-1 & 3.963E-1
\end{tabularx}
\caption{Time to reach the peak of non-axisymmetric magnetic energy and the peak value of that energy for the kinematic calculations in Fig.~\ref{kinematic}.}\label{scale}
\end{table}

\section{Selfconsistent MHD flow}\label{sec:mhd}

We now study the fully nonlinear MHD flow by numerically solving both (\ref{ns}) and (\ref{induction}). The importance of the magnetic forces can be estimated by comparing the peak magnetic energy \eqref{eq:Epeak} predicted kinematically to the energy available in the shear flow: this ratio is $\sim \Va^2 \Rm^{2/3}$ for $m=1$. Thus if $\Va\gtrsim \Rm^{-1/3}$, the non-axisymmetric field can be expected to act as a brake on the large-scale shear before it is annihilated by diffusion (\citealt{Spruit1999}).
For $\Re^{-1}\ll \Va\ll \Rm^{-1/3}$, the field will be symmetrized but will eventually drive the flow toward isorotation (in the sense of Ferraro's law) before viscosity enforces solid-body rotation. Finally, if $\Va\ll \Re^{-1}$, the kinematic approximation should describe the flow reasonably well at all times. All this presumes that the axisymmetric and non-axisymmetric field strengths are initially comparable. The effect of
varying the misalignment angle $\alpha$ is described at the end of this section.
Unfortunately, numerical considerations dictate that we adopt $\Re=\Rm$ rather than $\Re\gg \Rm$, so we must use some care to distinguish magnetic from viscous effects.

The influence of the non-axisymmetric field can also be described as a form of phase mixing, but one that involves azimuthal advection by the velocity field as well as Alfv\'enic propagation along the lines.  As in the axisymmetric case, fluid elements on the same field line can exchange angular momentum through magnetic tension, and differing rates of propagation on neighboring lines leads to phase mixing and damping of the associated Alfv\'en waves.  But since the field is non-axisymmetric, the projections of magnetic field lines onto the meridional may intersect.  Insofar as the velocity field remains approximately axisymmetric and predominantly rotational, the differential rotation may bring points on two such lines arbitrarily close together, even though these points are initally widely separated in azimuth.  The combination of these Alfv\'enic and advective processes may drive the entire volume to a uniform angular velocity, provided that the non-axisymmetric field persists.

\begin{figure}[h]
\centering
\includegraphics[scale=0.8]{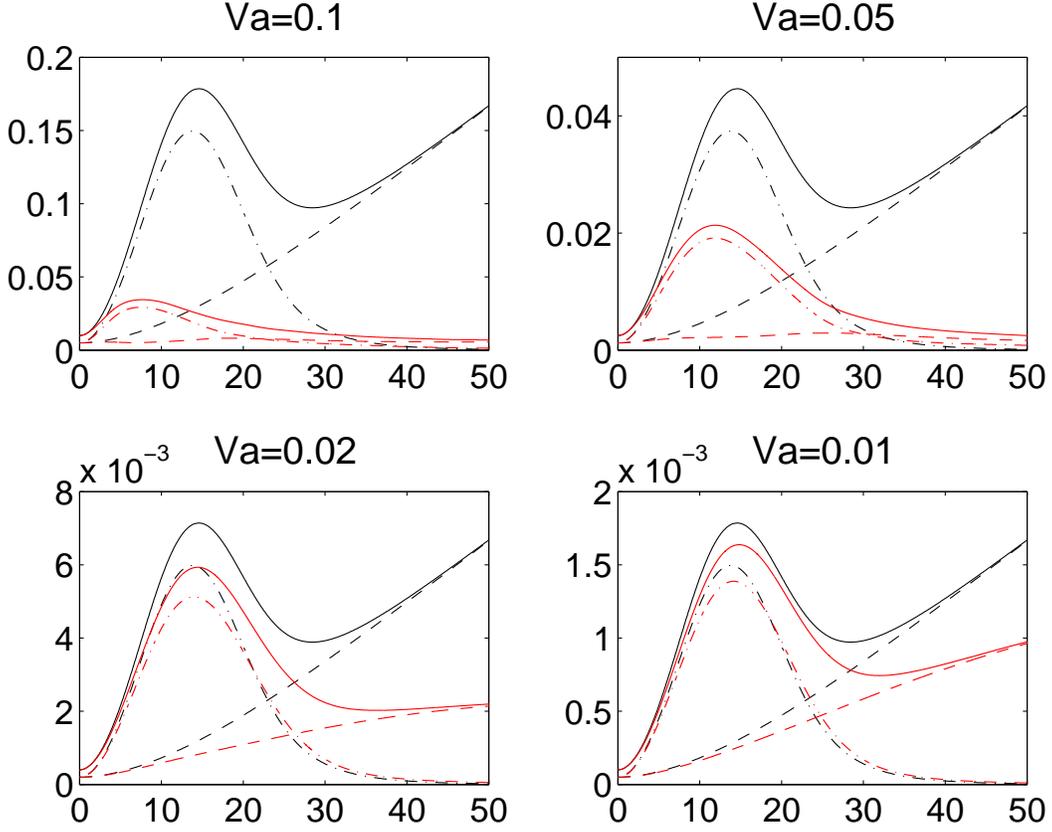}
\caption{Time evolution of total (solid), axisymmetric (dashed) and non-axisymmetric (dash-dot) magnetic energy in both kinematic problem (black) and selfconsistent flow (red). $\Re=\Rm=5000$ and $\alpha=45^\circ$.}\label{kinematic-mhd}
\end{figure}

\begin{figure}[h]
\centering
\subfigure[]{\includegraphics[scale=0.4]{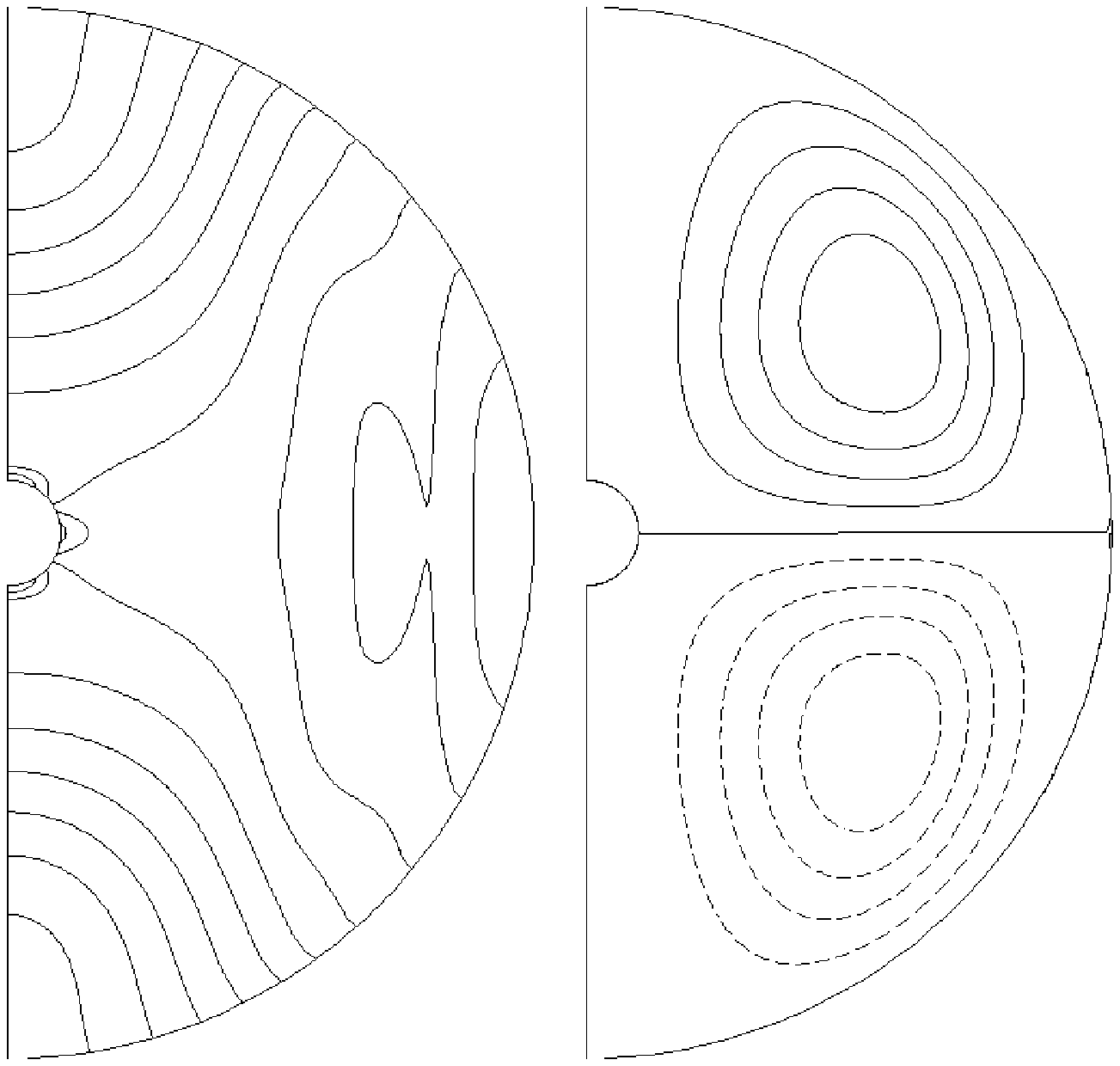}\label{u}}
\subfigure[]{\includegraphics[scale=0.7]{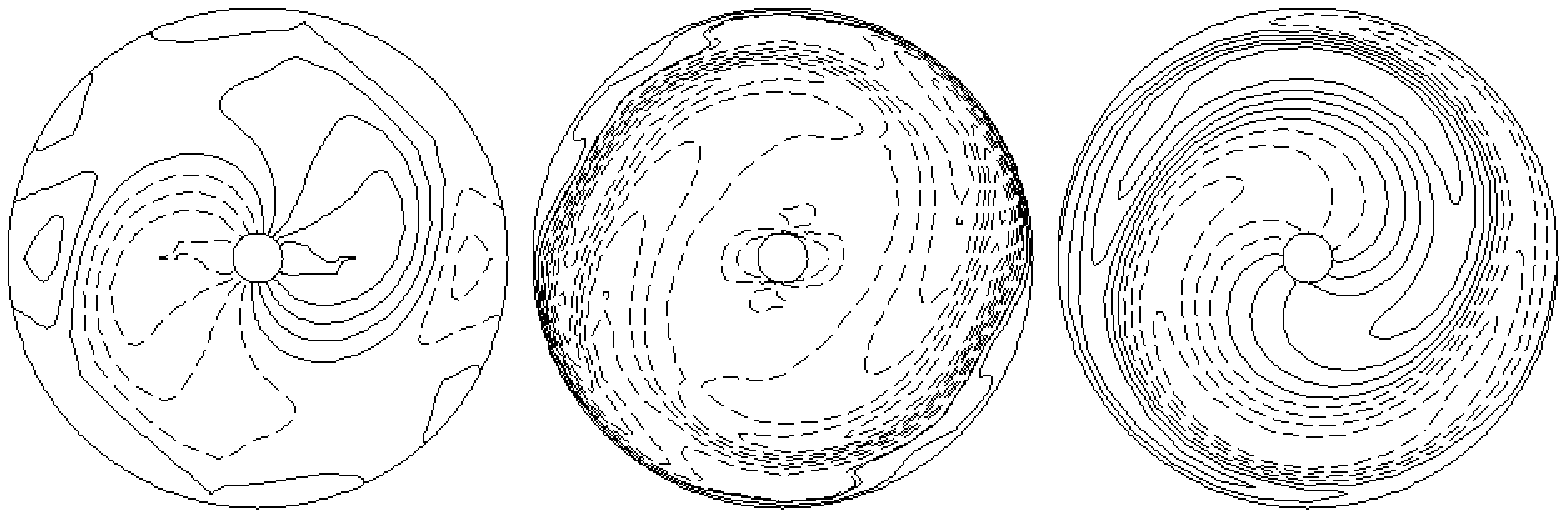}\label{b}}
\caption{Sequel to Fig.~\ref{initial} at the time when the non-axisymmetric magnetic energy peaks ($t=7.48$);
 $\Re=\Rm=5000$, $\Va=0.1$ and $\alpha=45^\circ$.
(a) Contours of axisymmetric angular velocity (left panel) and meridional circulation (right panel) in the meridional plane.
(b) Contours of $B_r$, $B_\theta$ and $B_\phi$ in the equatorial plane. In both (a) and (b), solid lines denote positive and dashed negative.}\label{u-b}
\end{figure}

First we study the magnetic back reaction by varying the initial field strength. Figure~\ref{kinematic-mhd} compares the time evolution of magnetic energy in the kinematic problem with that of the selfconsistent MHD flow at different $\Va$.  Evidently, the latter approaches the kinematic behavior at low $\Va$. In the first panel, the magnetic field is so strong ($\Va=0.1>\Rm^{-1/3}\approx0.06$) that it modifies the rotation profile significantly, as can be seen in Figure~\ref{u}.  The peak magnetic energy of the selfconsistent solutions is limited by the excess kinetic energy available in the initial differential rotation ($\Delta E_0\approx 0.0455$, \S\ref{sec:formulation}).

Fig.~\ref{u-b} shows the flow and field for the same simulation as in the first panel of Fig.~\ref{kinematic-mhd} at the time when the non-axisymmetric magnetic energy peaks, $t=7.48$.  The first panel shows a large change in the pattern of differential rotation compared to the initial state (Fig.~\ref{b0-omega0}).  Meridional circulation is induced, as shown in the right panel.  The winding up of the field is evident in the plan views shown in the last three panels.

\begin{figure}[h]
\centering
\subfigure[]{\includegraphics[scale=0.4]{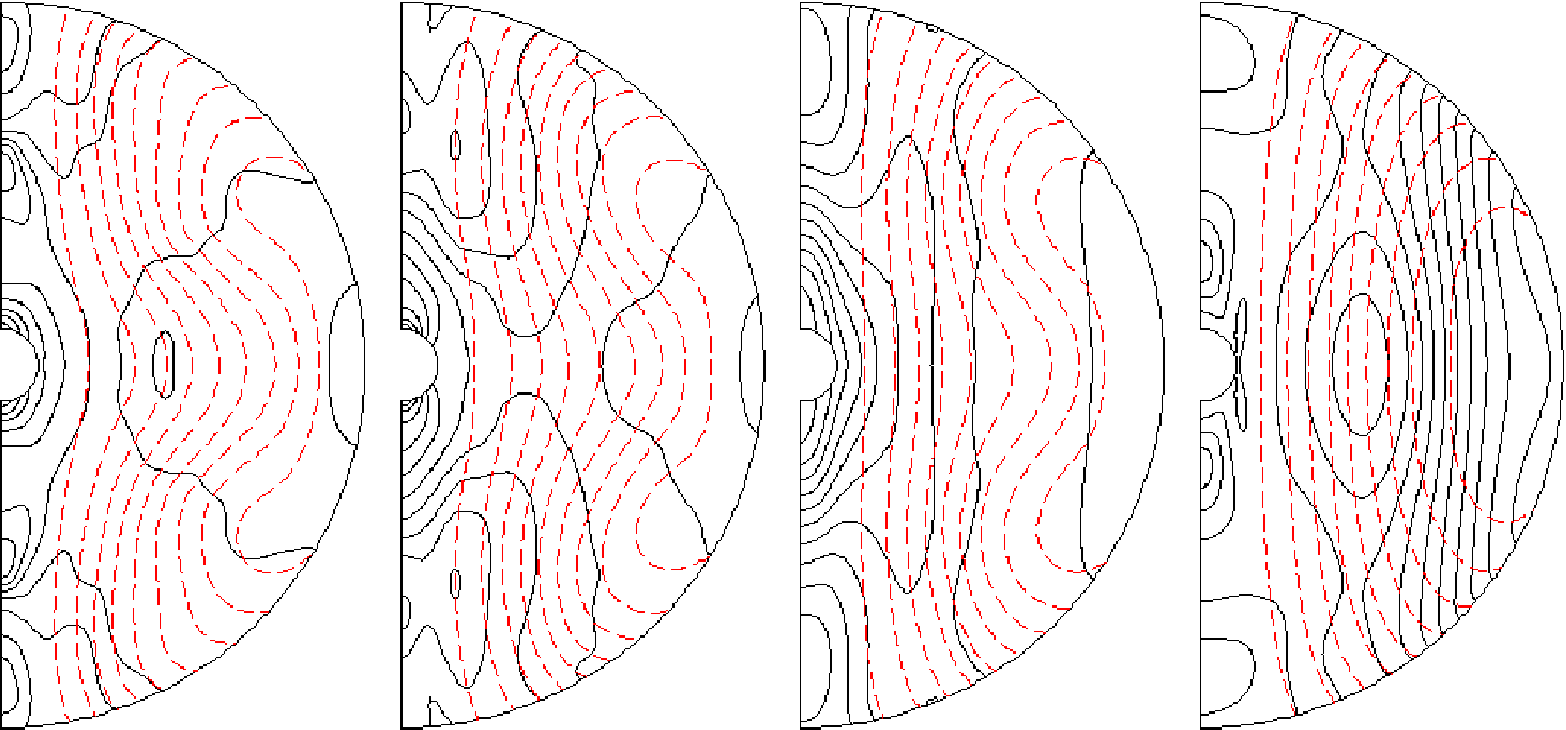}\label{b-omega-va0.1}}
\subfigure[]{\includegraphics[scale=0.4]{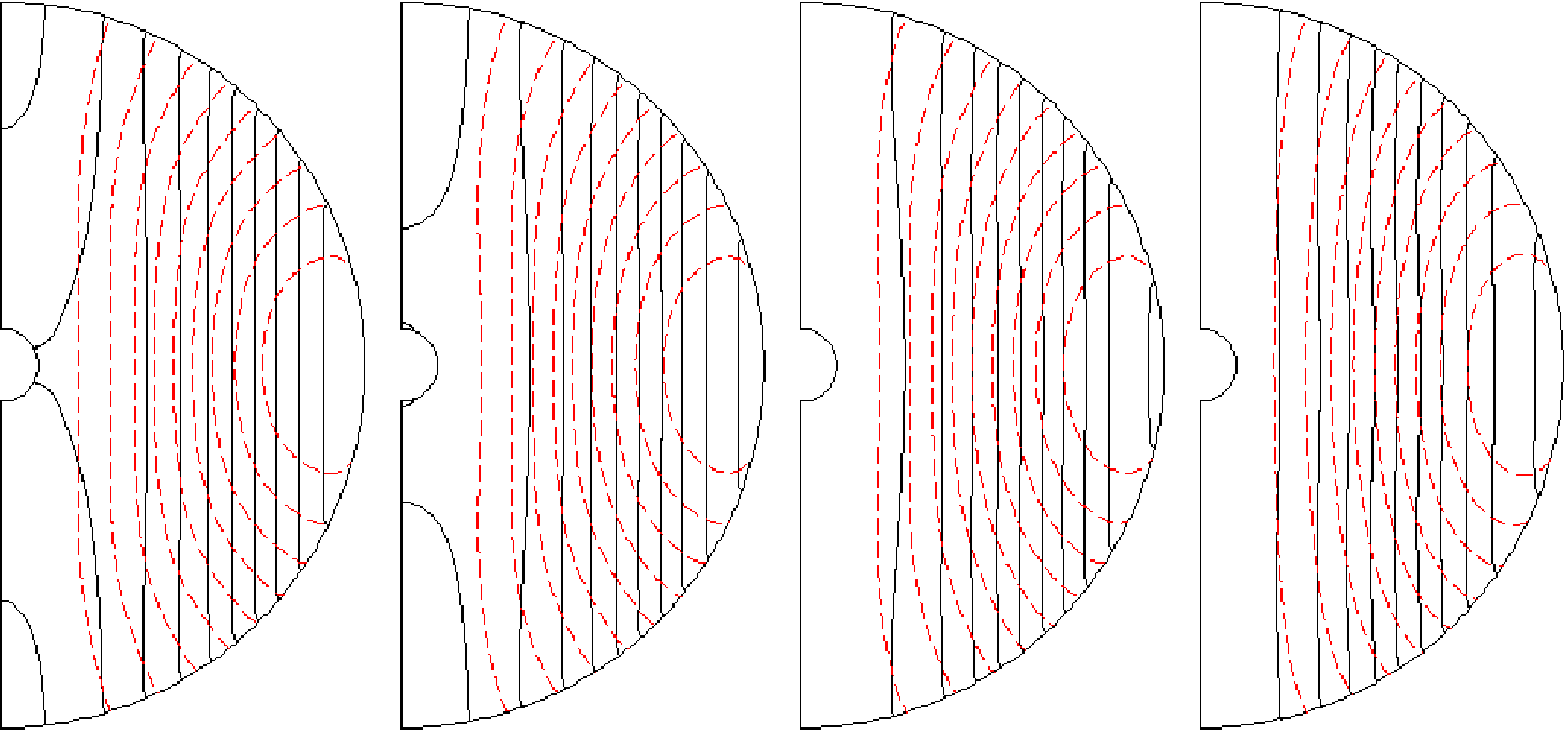}\label{b-omega-va0.01}}
\caption{Contours of axisymmetric angular velocity (black lines) and magnetic field (red lines) in the meridional plane. $\Re=\Rm=5000$ and $\alpha=45^\circ$. (a) $\Va=0.1$, (b) $\Va=0.01$. In both (a) and (b) the four panels from left to right are at time $=30$, $40$, $50$ and $100$.}\label{b-omega}
\end{figure}

Next we discuss the applicability of the isorotation law $\bm B_p\cdot\bm\nabla\Omega=0$.  Non-axisymmetric field alters the conditions for isorotation. Suppose that $\bm u=R\Omega\hat{\bm e}_\phi$ and $\bm B=B_R\hat{\bm e}_R+B_z\hat{\bm e}_z$ in cylindrical coordinates $(R,\phi,z)$, and that $\Rm\rightarrow\infty$ so that magnetic diffusion  is ineffective.  In axisymmetry, the induction term
$\bm\nabla\times(\bm u\times\bm B)=R\left(\bm B_p\cdot\bm\nabla\Omega\right)\hat{\bm e}_\phi$.  If this vanishes, then
$\partial\bm B/\partial t=\bm0$, so that isorotation can be maintained indefinitely, with different angular velocities on different
magnetic surfaces (surfaces parallel to the field lines).
When the field depends nontrivially on the azimuthal coordinate ($\phi$), however, the induction term becomes
\begin{equation*}
\bm\nabla\times(\bm u\times\bm B)=R\left(\bm B_p\cdot\bm\nabla\Omega\right)\hat{\bm e}_\phi-\Omega\left(\frac{\partial B_R}{\partial\phi}\hat{\bm e}_R+\frac{\partial B_z}{\partial\phi}\hat{\bm e}_z\right).
\end{equation*}
Even if $\bm B_p\cdot\bm\nabla\Omega$ is initially zero, the second term on the right side will cause $\bm B_p$ to evolve, so that additional constraints must be satisfied by the initial state to maintain isorotation.  It is likely that there are very few if any nonaxisymmetric isorotational states other than solid-body rotation \citep{Mestel_Weiss}.  Since the non-axisymmetric field decays much faster than the axisymmetric field, however, it may be that isorotation can be approached on intermediate timescales when the field is substantially axisymmetric but viscosity has not yet eliminated all shear in the velocities.  Such intermediate states are possible in the regime $\Re^{-1}\ll \Va\ll \Rm^{-1/3}$. Figure \ref{b-omega} shows the contours of axisymmetric angular velocity and poloidal field lines at $\Va=0.1$ and $0.01$: the former is outside the regime in question, while the latter is within it. In each subfigure the four panels are taken at intermediate times after the peak of the non-axisymmetric magnetic energy (see upper-left and bottom-right panels in figure \ref{kinematic-mhd}) but much earlier than $\Omega^{-1}\Re$. Figure \ref{b-omega-va0.1} suggests that the isorotational state is destroyed by the stronger field, while Fig.~\ref{b-omega-va0.01} suggests that it is nearly achieved for the weaker field in the deeper interior.  In the latter case, the angular velocity is nearly a function of cylindrical radius alone, as required for steady unstratified flow when magnetic and other non-potential forces are unimportant.  (Some dependence on $z$ as well as $R$ might have resulted if we had allowed for stratification.)  The magnetic field lines are approximately parallel to the rotation axis at depth but not near the surface.  They resemble the shape of the slowest-decaying resistive magnetic eigenfunction in a uniformly rotating and uniformly conducting shell, especially near the equator (Fig~\ref{b0-omega0}).

Thirdly we compare the relative strength of differential rotation and misaligned field. As pointed out by \citet[ \S9.3]{Mestel}, the degree of misalignment is crucial to the dynamics. To quantify the relative strength we compare two characteristic times. The first is the time when the excess kinetic energy drops to 10\% of its initial value.  As noted in \S2, the kinetic energy decays from $0.3191$ initially to $0.2736$ in a state of uniform rotation at the same total angular momentum; thus the reduction of the difference of these by 90\% corresponds to kinetic energy $=0.2781$. The second characteristic time is that when the non-axisymmetric magnetic energy drops to 10\% of its peak (not initial) value. Table~\ref{time} lists these two times for several $\Re=\Rm$ and $\Va$.  On each column higher $\Va$ corresponds to faster decay of kinetic energy and slower decay of non-axisymmetric magnetic energy.  A stronger field tends to convert more kinetic energy to magnetic energy, as already seen in Fig.~\ref{kinematic-mhd} and the surrounding discussion above. More importantly, in the regime $\Va\le0.03$ the time for the 90\% drop of kinetic energy is later than the time for the 90\% drop of non-axisymmetric magnetic energy, whereas in the regime $\Va\ge0.05$ the situation is reversed. Therefore, there exists a boundary in the $(\Rm,\Va)$ plane across which the dynamics is qualitatively changed. Below this boundary the magnetic field is symmetrized before differential rotation completely decays; above it, differential rotation is suppressed before the magnetic field is symmetrized (if it is symmetrized at all before it decays). This is consistent with the arguments of \cite{Mestel} and \cite{Spruit1999}.

To test these arguments more quantitatively, we have interpolated in $\Va$ along each column of Table~\ref{time} to find the value $\Va_{\rm crit}$ at which the two times in question are equal.  The bottom row of Table~\ref{time} indicates that $\Va_{\rm crit}$ is not a monotonic function of $\Rm$.  A power-law fit to the final two colums (the highest $\Rm$) yields $\Va_{\rm crit}\propto\Rm^{-0.20}$, a somewhat weaker dependence than the scaling $\Rm^{-1/3}$ predicted by the quasi-kinematic reasoning at the beginning of this section.  We return to this point in \S\ref{sec:discussion} below.

\begin{table}
\noindent
\begin{tabularx}{\textwidth}{X|X|X|X|X|X}
 & 500 & 1000 & 2000 & 5000 & 10000 \\
\hline 
0.01 & 20.6/17.9 & 39.5/20.2 & 75.0/23.9 & 159.4/31.1 & 258.8/38.2 \\
\hline
0.02 & 19.8/18.1 & 36.5/20.5 & 64.1/24.2 & 115.6/31.6 & 185.9/38.9 \\
\hline
0.03 & 18.4/18.4 & 32.0/21.0 & 50.7/25.1 &  72.5/32.3 &  60.4/38.7 \\
\hline
0.04 & 16.5/18.9 & 26.3/21.7 & 35.3/26.0 &  34.2/32.7 &  30.2/45.0 \\
\hline
0.05 & 14.4/19.5 & 20.0/22.9 & 21.8/27.0 &  23.9/33.4 &  25.4/47.0 \\
\hline
0.1  &  8.5/21.3 &  9.8/26.2 & 11.8/31.0 &  19.3/37.0 & --- \\
\hline
 & 0.0300 & 0.0461 & 0.0464 & 0.0414 & 0.0359
\end{tabularx}
\caption{The time when the kinetic energy drops 90\% of its initial value (on the left of slash) and the time when the non-axisymmetric magnetic energy drops 90\% of its peak value (on the right of slash). The top row indicates the values of $\Re=\Rm$, the left column $\Va$ and the bottom row $\Va_{\rm crit}$. $\alpha=45^\circ$. The calculation at $\Re=\Rm=10000$ and $\Va=0.1$ was incompletely resolved.}\label{time}
\end{table}

\begin{figure}[h]
\centering
\includegraphics[scale=0.8]{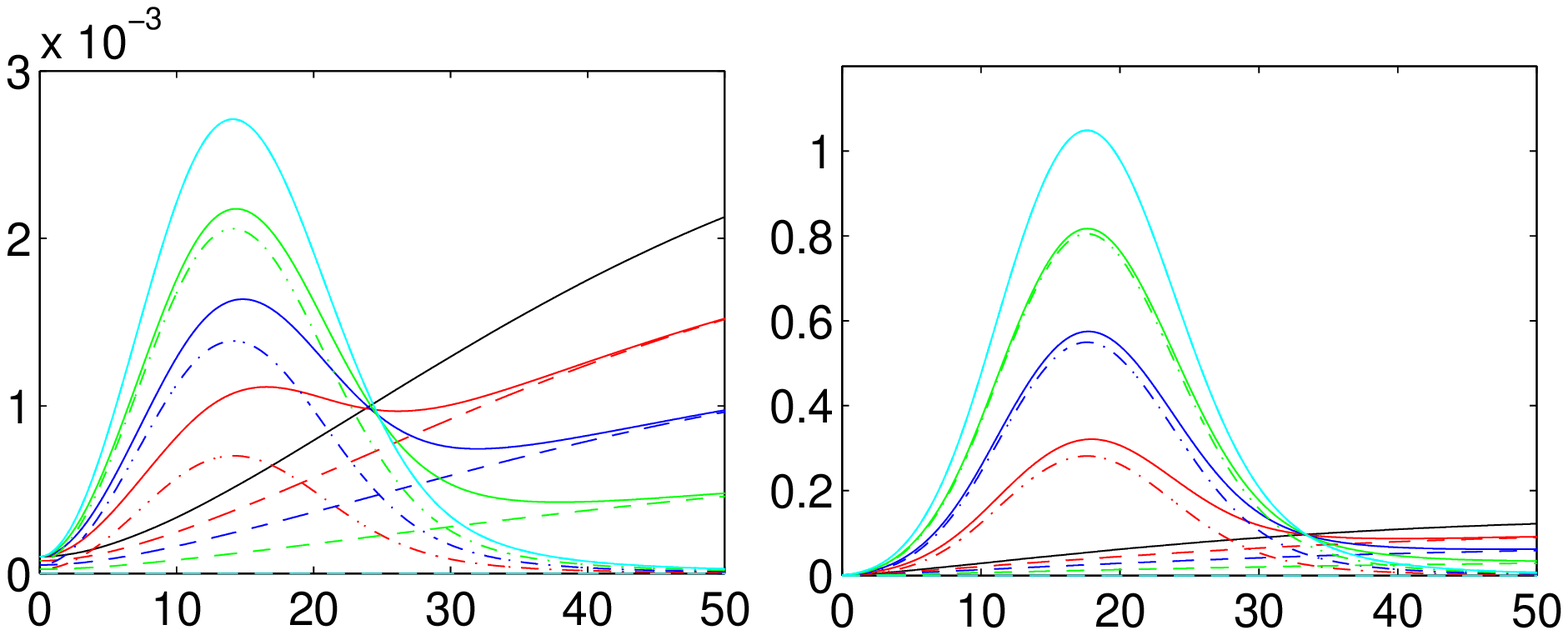}
\caption{Time evolution of magnetic energy (left panel) and Ohmic dissipation (right panel) at different magnetic obliquities.
$\Re=\Rm=5000$ and $\Va=0.01$. Black, red, blue, green and cyan lines denote respectively $\alpha=0^\circ$, $30^\circ$, $45^\circ$, $60^\circ$ and $90^\circ$. Solid, dashed and dash-dot lines denote respectively the total, axisymmetric and non-axisymmetric energy or dissipation ($\alpha=0^\circ$ and $\alpha=90^\circ$ are entirely axisymmetric and non-axisymmetric, respectively).}\label{angle}
\end{figure}

To end this section, we briefly discuss the effect of varying the angle $\alpha$ between the rotational and magnetic axes. Since, as we have seen, there are strong differences between the axisymmetric cases (where $\alpha=0$) and those for which $\alpha=45^\circ$, it is reasonable to expect that the strength of the interaction between the flow and the field should increase continuously with this angle up to $\alpha=90^\circ$. This expectation is tested in Figure~\ref{angle}, which shows the evolution of the magnetic energy and Ohmic dissipation for several values of $\alpha$. A larger angle leads to higher energy and dissipation, as expected. In fact, for the cases shown in Fig.~\ref{angle} and tabulated in Table~\ref{table-angle}, the peak of the non-axisymmetric energy increases linearly with $\sin^2\alpha$ (correlation coefficient $0.9999$).  This would not be surprising in the kinematic problem, since the initial amplitude of the non-axisymmetric component is proportional to $\sin\alpha$, but the calculations shown in Fig.~\ref{angle} include magnetic backreaction. Of course the energy $E_{\rm mag,peak}$ (and somewhat more, because dissipation has already begun to act at the time of the peak) must come at the expense of the differential rotation.

\begin{table}
\begin{tabularx}{\textwidth}{X|X|X|X|X|X}
$\alpha$ & $0^\circ$ & $30^\circ$ & $45^\circ$ & $60^\circ$ & $90^\circ$ \\
\hline
$E_{\rm mag,\,peak}$ & 0 & 7.019E-4 & 1.387E-3 & 2.057E-3 & 2.711E-3
\end{tabularx}
\caption{Peak non-axisymmetric magnetic energy versus magnetic obliquity $\alpha$, following Fig~\ref{angle}. $\Re=\Rm=5000$ and $\Va=0.01$.}\label{table-angle}
\end{table}

\section{Discussion}\label{sec:discussion}

In this work we have explored numerically the interaction of differential rotation and a misaligned magnetic field. In the kinematic limit we verify the $\Rm^{1/3}$ law for the time and amplitude at which the non-axisymmetric azimuthal field reaches its peak and then decays, whereas the corresponding scaling for the axisymmetric field is $O(\Rm)$.  In selfconsistent calculations where the flow reacts to magnetic forces, a sufficiently weak magnetic field behaves approximately as in the kinematic problem, becoming axisymmetric before the rotation becomes uniform, and a nontrivial state of isorotation---one in which different magnetic surfaces have different angular velocities----is approximately achieved near the rotation axis.  For stronger fields, there exists a boundary in the plane of dimensionless initial magnetic field strength $\Va$ versus $\Rm$ above which differential rotation is suppressed before the nonaxisymmetric magnetic field decays.  This boundary is better described by $\Va_{\rm crit}\propto\Rm^{-0.20}$ than by the expected scaling $\Rm^{-1/3}$, at least over the range explored by our simulations ($\Rm\le 10^4$). We do not understand this quantitatively for our actual simulations, though according to the arguments given below, one expects the dependence of $\Va_{\rm crit}$ on $\Rm$ to weaken at very large $\Rm$ because of reconnection and perhaps other nonlinear processes that assist in the destruction of the nonaxisymmetric field.  We have also verified that a larger misalignment angle ($\alpha$) between the rotational and magnetic axes leads to stronger exchanges between the flow and the field, in proportion to $\sin^2\alpha$ (Fig.~\ref{angle}).

In our calculations the magnetic Prandtl number $\Pm\equiv\Rm/\Re=\nu/\eta$ has been fixed at unity for numerical
  convenience. With the code and computational resources available to us, fully resolved simulations are practical only
  if the larger of $\Re$ and $\Rm$ is $\lesssim 10^4$. Therefore, we approach ideal MHD best by taking $\Re=\Rm$. Direct
  numerical simulations of magnetic dynamos \citep[e.g.][]{Schekochihin+etal2007} and magnetorotational turbulence
  \citep[e.g.]{Fromang+etal2007} often find that $\Pm$ influences large-scale properties of the flow, at least within
  the computationally accessible range of $\Re$ and $\Rm$. Whether $\Pm$ remains macroscopically important at much
  larger $\Rm$ and $\Re$, as in real stars and in the interstellar medium, can only be addressed at present via
  idealized models of turbulence (\citealt{Kulsrud+Anderson1992,Malyshkin+Boldyrev2009,Schober+etal2012}). It is worth
  noting, however, that in comparison to $\Re$ and $\Rm$ themselves, $\Pm$ is not so far from unity in the applications
  of interest to us. At $r=0.5\Rsun$ in the Sun (where $T\approx4\times 10^6\mathrm{K}$ and
  $\rho\approx 1.3\mathrm{g\,cm^{-3}}$), one estimates from the classical Spitzer formulae that $\Pm\approx 10^{-2}$.
  For white-dwarf interiors at $T\sim 10^6$-$10^7\mathrm{K}$ and $\rho\sim10^6\mathrm{g\,cm^{-3}}$, we
  estimate\footnote{using the results of \citet{Yakovlev+Urpin1980} as encoded in the \textsc{fortran} programs at
    \url{www.ioffe.ru/astro/conduct}, which return electrical and thermal conductivities. We roughly translate the
    latter to a kinematic viscosity by multiplying by $m_i/k_{\textsc{b}}\rho$, $m_i$ being the ion mass, presuming that
    the material has not crystallized.} $\Pm\sim 10^2$-$10^3$.

Another simplification we have made is to neglect stratification, an important effect in the outer parts of white dwarfs,
where there are significant entropy and composition gradients, and in the radiative zone of the Sun.  As noted in \S1,
stratification would allow a larger range of isorotational final states in hydrostatic equilibrium.  The ratio of Brunt-V\"as\"al\"a
frequency to rotation frequency ($N/\Omega$) is important for the growth rates of magnetic instabilities that may gradually
modify the differential rotation as well as the field itself \citep{Spruit1999}.  However, we do not expect that stratification has
a major effect on the interaction between a non-axisymmetric field and differential rotation (e.g., the dependence of $\Va_{\rm crit}$ on $\Rm$), at least on the dynamical timescales simulated here.  Nor have we allowed for compressibility or density gradients.
All of these could be the subject of future work, perhaps based on realistic models of representative stellar models.

Notwithstanding these caveats, we tentatively suggest some implications for real stars.
As discussed in \S1, this work has been motivated in part by tidal interactions between white dwarfs in binary orbits decaying under the influence of gravitational radiation.  The calculations we have performed, however, are not directly applicable to that astrophysical problem. For one thing, the magnetic Reynolds number of a white dwarf is enormous: $\Rm\sim 10^{17}$ for diffusivity $\eta\approx 1\unit{cm^2\,s^{-1}}$ \citep{Yakovlev+Urpin1980}, radius $R\approx0.01 \unit{R_\odot}\approx7\times10^8\unit{cm}$, and angular velocity $\Omega=2\pi/1\unit{min}$.  Solar flares and other more theoretical evidence suggest that under such nearly ideal conditions, the destruction of magnetic energy proceeds more rapidly than by linear resistive diffusion.  It is plausible that nonlinear dissipative processes are already important at the modest values of $Re$ and $\Rm\sim 10^3$ in our simulations.  This may account for the difference between the predicted and observed scalings of $\Va_{\rm crit}$ with $\Rm$.  While a variety of collisionless-plasma effects have been invoked to explain the observed rate magnetic energy release in solar flares, recent theoretical work indicates that even in classical resistive MHD, oppositely-directed field lines may approach one another and annihilate at a speed $V_{\rm rec}\approx f\VA$ that is independent of the true microscopic diffusivity ($\eta$), provided that the Lundquist number $S\equiv L\VA/\eta\gtrsim 10^4$, where $\VA$ is the physical Alfv\'en speed and $L$ is an appropriate macroscopic lengthscale, such as the length of the reconnecting current sheet; the present estimate for the dimensionless factor is $f\approx 0.02$ \citep[and references therein]{Loureiro+etal2012}.  Mechanisms by which $V_{\rm rec}/\VA$ becomes independent of $\eta$ (or perhaps depends upon it only very weakly) are said to provide ``fast reconnection''.

In our problem, $\VA$ should be based on the tightly wound azimuthal field, which scales with time $t$ as $\VA\sim \Va\, R\Omega \Delta\Omega t$ as explained in \S\ref{sec:kinematic}, while the lengthscale along field lines is $L\sim r_0$.  Thus $S\sim \Va\,\Rm'$.  The local timescale on which the field is destroyed now becomes $l/V_{\rm rec}$ rather than $l^2/\eta$, with $l\sim r_0/(t\Delta\Omega)$ being the cross-field lengthscale on which the field reverses.  Equating this local timescale to the winding time $t$ leads to a revised estimate of the time at which the magnetic energy should reach its peak: $\Delta\Omega t_{\rm pk}\sim(f\Va\Omega/\Delta\Omega)^{-1/3}$ rather than $\Rm^{-1/3}$ as before.  Defining $\Va_{\rm crit}$ so that the magnetic energy at the peak is equal to the excess kinetic energy $\Delta E$ in the initial differential rotation leads to $\Va_{\rm crit}\sim (\Omega/f\Delta\Omega)^{1/2}(\Delta E/5 E_\infty)^2$, where $E_\infty=L_z^2/2I$ is the final rotational energy, and the factor $1/5$ arises from $I/Mr_0^2\approx 2/5$ as for a full sphere of mass $M$ and constant density.  Finally, setting $f\approx 0.02$, $\Delta\Omega/\Omega\approx 1$, and $\Delta E/E_\infty\approx 1/6$ yields $\Va_{\rm crit}\approx 0.011$.  In short, if we were able to extend our calculations to $\Re=\Rm\to\infty$ with the same initial rotation profile and magnetic geometry, then we would expect $\Va_{\rm crit}$ to asymptote to $\sim 10^{-2}$ because of fast reconnection.  It is unlikely that our simulations achieve fast reconnection, however, because the Lundquist number at the peak is
\begin{equation*}
  S_{\rm pk}\sim \Rm\left(\frac{\Va^2\Delta\Omega}{f\Omega}\right)^{1/3}\sim 0.4\Rm
\end{equation*}
where we have put $Va=0.04$ and $\Delta\Omega/\Omega =1$ in the final estimate. Thus our largest-$\Rm$ simulations
probably undergo stable Sweet-Parker reconnection, for which $V_{\rm rec}\propto\VA^{1/2}$ and an argument along the
lines above predicts $\Va_{\rm crit}\propto\Rm^{-1/4}$. In astrophysical applications where $S\gg 10^4$, it is likely
that fast reconnection dominates so that $\Va_{\rm crit}$ becomes independent of $\Rm$.

In an effort to cast doubt on contemporary helioseismological evidence for differential rotation in the solar
  core, \citet{Mestel_Weiss} estimated that a meridional field $B_{\textsc{p}}>0.03\mathrm{G}$ would be sufficient to
  establish and maintain isorotation in the radiative zone. They admitted that isorotation does not require uniform
  rotation but went on to speculate that the latter would result if the magnetic field were ``even slightly
  non-axisymmetric.'' We are perhaps in a position to quantify this. From the results of Fig.~\ref{angle} and
  surrounding discussion, Mestel and Weiss's estimate (of conditions sufficient to establish uniform rotation) can be
  sharpened to $B_{\textsc{p}}\sin\alpha >0.03\mathrm{G}$, where $\alpha$ is the magnetic obliquity. But this presumes
  that the differential rotation does not symmetrize the field (reduce $\alpha$ to zero) before solid-body rotation is
  established. If the field in the core is frequently regenerated by a dynamo process, we know of no way to constrain
  its obliquity. But if it is a fossil field established in the early history of the Sun, then the free-decay problem we
  have studied may apply. The magnetic Reynolds number of the radiative zone is
  $\Rm_\odot\equiv R_c^2\Omega_\odot/\eta\sim 10^{14}$, where we take $R_c\approx0.7\Rsun$ for the outer radius of this
  zone and evaluate $\eta$ for $T=4\times10^6\mathrm{K}$ and $\rho=1.3\mathrm{g\,cm^{-3}}$ (i.e., conditions at
  $0.5\Rsun$). If $\Va_{\rm crit}=\Rm^{-1/3}$, then to avoid symmetrization, the field and obliquity would have to
  satisfy $B_{\textsc p}\gtrsim 10\mathrm{G}$, considerably stronger than the estimate above but still quite modest.
  However, the Lundquist number $R_c\Omega_\odot/\eta$ of the radiative zone is on the order of
  $10^8 (B/\mathrm{1\,G})$, so according to the discussion above, we expect to be in the regime of fast reconnection
  where $\Va_{\rm crit}$ becomes independent of $\Rm$. Adjusting moment of inertia in the argument above for the actual
  density profile of the solar core \citep{Bahcall+Pinsonneault1995}, we estimate $\Va_{\rm crit}\approx0.004$, which
  corresponds to $B_{\textsc{p}}\sin\alpha\approx 2\mathrm{kG}$. It is still debated whether a fossil magnetic field in the solar radiative core can enforce solid-body rotation, however.  A crucial issue is whether such a field can remain closed
within the core, because if it were to connect to the convection zone, then by Ferraro's Law the core also should
rotate differentially 
\citep{Gough+McIntyre1998,MacGregor+Charbonneau1999,Brun+Zahn2006,Garaud+Garaud2008,Acevedo+etal2013}.


A second difference between our simulations and the binary-white-dwarf problem is that
we have studied the free decay of a pre-existing profile of differential rotation rather than gradual acceleration by a
tidal torque.  (This allowed the outcome to be discerned with less CPU time.)  In the tidal-binary scenario, the system
starts with an orbital period of a few hours, which is short enough to lead to merging of a pair of $0.7\, M_\odot$
white dwarfs within $10^9\unit{yr}$ (\S\ref{sec:intro}).  Assuming that the tidal torque is absorbed in the outer layers
of the star, the magnetic stress necessary to maintain synchronous rotation of the interior is
\begin{equation*}
B_r B_\phi \sim \frac{I}{R^3}\frac{d\Omega}{dt} \sim  1\left(\frac{P}{1\unit{min}}\right)^{-11/3}\unit{MG^2},
\end{equation*}
in which $I\approx 0.2 MR^2$ is the moment of inertia of a cool white-dwarf model at this mass. Thus for example, if the
meridional field is $\sim 10\unit{kG}$, a plausible upper limit for most white dwarfs, then only a slight bending of the
lines is needed to maintain synchronism when the period is an hour or more. When the period is about a minute, however,
$B_\phi\sim 10^4 B_r\sim 100\unit{MG}$ would be required since differential rotation alone will not increase $B_r$.
While this is not outside the range of surface fields observed in some white dwarfs, a \emph{non-axisymmetric} field so
tightly wound would have a resistive time $\sim 10^{-8}$ times smaller than that of the most slowly decaying magnetic
eigenmode ($\sim 3\times 10^9\unit{yr}$); and magnetic reconnection would probably act on even shorter timescales, as
discussed above. Hence an axisymmetric field might be expected, unless perhaps dissipation in the interior leads to
nonaxisymmetric turbulence that produces a magnetic dynamo. In fact, as \citet{Spruit2002} has emphasized, a
predominantly azimuthal \emph{axisymmetric} field should be subject to nonaxisymmetric Tayler instabilities, though in
strongly stably-stratified regions these instabilities will be concentrated toward the poles, whereas it is near the
equator that meridional field is most needed to transmit tidal torques. Also, while some numerical evidence
for such a dynamo has been claimed by \citet{Braithwaite}, it is still debated whether there exists dynamo action in stably
stratified zones (e.g., \citealt{Zahn+Brun+Mathis2007}).  If it exists, such a dynamo might maintain $B_r$ in
a constant ratio to $B_\phi$, allowing a lower overall stress for the same torque. By our estimates, a radial field
of order $1\unit{MG}$ would also begin to have a significant effect on the dispersion relation of g-modes at the
compositional interface where they are tidally excited in the models of \cite{Fuller+Lai2012}.

In summary, while there is little doubt that the interactions among tidal torques, differential rotation, and magnetic
fields are important for coalescing white-dwarf binaries, we are far from being able to predict the tidal heating and
luminosity of such systems in the last stages of inspiral.

\acknowledgements We thank Prof. Rainer Hollerbach for help with the calculation of energy and dissipation in his spectral code, and an anonymous referee for questions and criticism that materially improved the presentation.  JG thanks the Australian National University and Mt. Stromlo observatory for their hospitality while our paper was being revised.  This work was supported by the National Science Foundation’s Center for Magnetic Self-Organization under grant PHY-0821899.

\bibliographystyle{apj}
\bibliography{paper}

\end{document}